\documentclass[aps,prl,showpacs,twocolumn,floatfix,superscriptaddress]{revtex4}
\usepackage{graphicx}
\usepackage{amsmath}

\begin{document}

\title{Diffusion, subdiffusion, and trapping  of active particles in heterogeneous media}

\date{\today}

 \author{Oleksandr Chepizhko}
\affiliation{Odessa National University, Department for Theoretical Physics, Dvoryanskaya 2, 65026 Odessa, Ukraine}
\affiliation{Universit{\'e} Nice Sophia Antipolis, Laboratoire J.A. Dieudonn{\'e}, UMR 7351  CNRS, Parc Valrose, F-06108 Nice Cedex 02, France}

\author{Fernando Peruani}
\email{Peruani@unice.fr}
\affiliation{Universit{\'e} Nice Sophia Antipolis, Laboratoire J.A. Dieudonn{\'e}, UMR 7351  CNRS, Parc Valrose, F-06108 Nice Cedex 02, France}

\begin{abstract}
We study the  transport properties of a system of active particles moving at constant speed in an heterogeneous two-dimensional  space. 
The spatial heterogeneity is modeled by a random distribution of obstacles, which the active particles avoid. Obstacle avoidance is characterized by the particle turning speed $\gamma$. 
%
%
%
We show, through simulations and analytical calculations, that the mean square displacement of particles exhibits two regimes as function of the density of obstacles $\rho_o$ and $\gamma$.
%
%
We find that at low values of  $\gamma$, particle motion is diffusive and characterized by a diffusion coefficient that displays a minimum at an intermediate obstacle density $\rho_o$. 
We  observe that in high obstacle density regions and for large $\gamma$ values, spontaneous trapping of active particles occurs. 
We show that such trapping leads to genuine subdiffusive motion of the active particles. 
%
%
We indicate how these findings  
can be used to fabricate a filter of active particles. 
\end{abstract}


\maketitle

Locomotion patterns are of prime importance for the  survival of most organisms at all scales, ranging from bacteria to birds, 
and often involving  complex processes that require energy consumption: i.e. the active motion of the organism~\cite{berg, okubo}. 
The characterization and study of these patterns have a long tradition~\cite{berg, okubo} and the 
experimental observation of subdiffusion, diffusion, and superdiffusion has motivated 
the development of powerful theoretical tools~\cite{metzler2000}. 
It is only in recent years that the (thermodynamical) non-equilibrium nature of these active patterns has been exploited, 
leading to the study of the so-called active particle systems~\cite{romanczuk2012}. 
Exciting non-equilibrium features have been reported in both, interacting as well as non-interacting active particle systems. 
For instance, large-scale collective motion and giant number fluctuations have been found in interacting active particle systems~\cite{vicsek2012, ramaswamy2010, ginelli2010, peruani2012}.  
In non-interacting active particle systems, the presence of active fluctuations leads to complex, non-equilibrium transients in the particle mean square displacement~\cite{peruani2007, golestanian2009}   and anomalous velocity distributions~\cite{romanczuk2011},   
and the lack of momentum conservation induces non-classical particle-wall interactions, which allows, for instance, the rectification of particle motion~\cite{galajda2007, wensink2008, wan2008, tailleur2009, radtke2012}.  

The study of active particle systems has  recently witnessed the emergence of a  promising new direction:  
the design and construction of biomimetic, artificial active particles. 
The directed driving is usually obtained by fabricating asymmetric particles that possess two distinct friction coefficients~\cite{kudrolli2008, deseigne2010, weber2013}, 
light absorption  coefficients~\cite{jiang2010, golestanian2012, theurkauff2012, palacci2013}, or catalytic properties~\cite{paxton2004, mano2005, rucker2007, howse2007, golestanian2007, golestanian2009} depending on whether energy injection is done through vibration, 
light emission, or chemical reaction, respectively. 
One of the most prominent features of these artificial active particles is that their motion is characterized by a diffusion coefficient  remarkably larger than the one obtained using symmetric particles~\cite{golestanian2009}. 

\begin{figure} 
\begin{center}
\resizebox{\columnwidth}{!} {\includegraphics{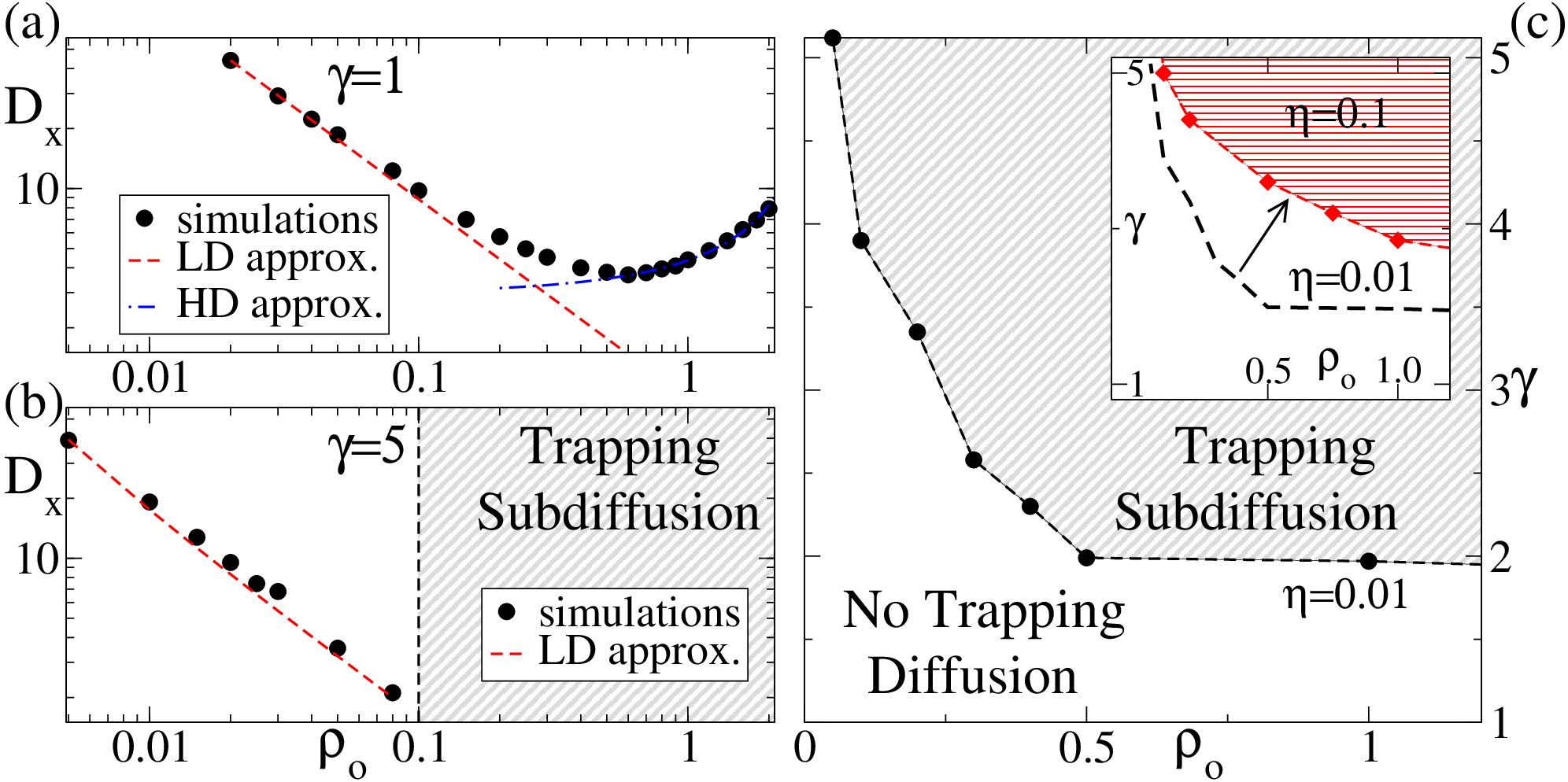}}
\caption{Diffusive and subdiffusive regimes. (a) For low values of the turning speed $\gamma$, the motion is diffusive and characterized by a diffusion coefficient $D_x$ that exhibits a minimum with the obstacle density 
$\rho_o$, as expected 
by combining the low-density (LD) and high-density (HD) approximation. 
%
See Eq.~(\ref{eq:rho_asymptotic}),~(\ref{eq:rho_asymptotic_HD}) and text. 
(b) For large values of $\gamma$, diffusive motion occurs at low $\rho_o$ values only, while for large $\rho_o$ values, particle motion becomes subdiffusive.  
(c) The boundary between the diffusive and subdiffusive regime is given,  for a fixed noise $\eta$, by $\rho_o$ and $\gamma$, see inset and~\cite{boundaryEstimate}.  
Parameters: $R=1$, $L=100$, $N_p=10^4$, and $\eta=0.01$ (inset in (c), $\eta=0.1$). 
 }
\label{fig:diffusion}
\end{center}
\end{figure}

The rapidly expanding study of active particles has focused so far almost exclusively, theoretically as well as experimentally, on the statistical description of particle motion in  idealized,  
homogeneous spaces.  
However, the great majority of natural active particle systems takes place, in the wild, in heterogeneous media:  
from active transport inside the cell, which occurs in a  space that is filled by organelles and vesicles~\cite{alberts}, 
to bacterial motion, which takes place in highly heterogeneous environments such as the soil or  complex tissues such as in the gastrointestinal tract~\cite{dworkin}.   
While diffusion in random media is a well studied subject~\cite{bouchaud1990, berry2011}, the impact that an heterogeneous medium may have on the locomotion patterns of active particles remains poorly explored. 
%

\begin{figure*}
\begin{center}
\resizebox{16.5cm}{!}{\rotatebox{0}{\includegraphics{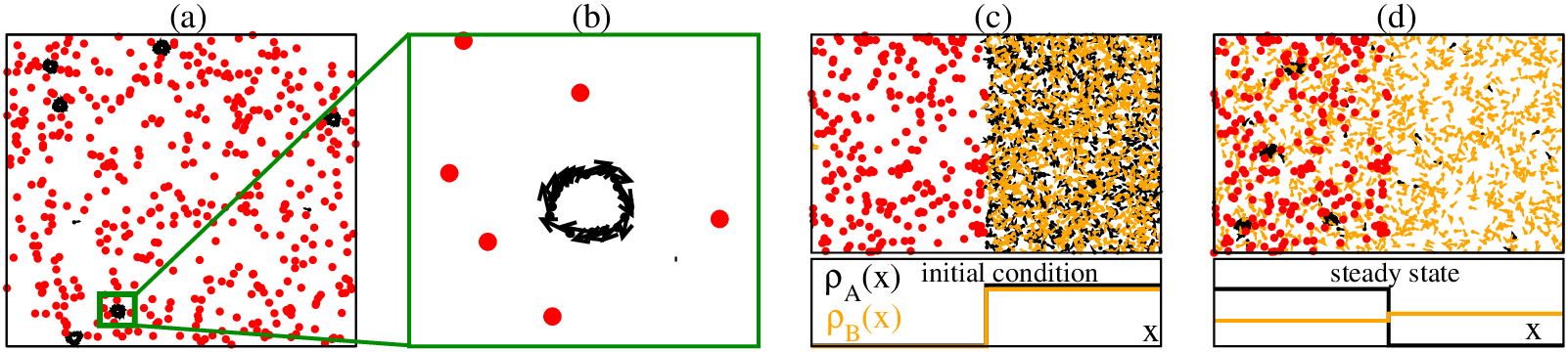}}}
\caption{Trapping and filtering.  For large values of $\gamma$ and $\rho_o$, spontaneous trapping of particle occurs, see (a) where $\gamma=5$,  $\rho_o=0.5$, $\eta=0.01$, and $L=30$. 
Obstacles are indicated by red dots while black arrows correspond to SPPs.  
Inside traps particles self-organize into vortex-like structures, (b)~\cite{commentMovies}.  
Spontaneous trapping leads to subdiffusion as indicated in Fig.~\ref{fig:subdiffusion} and can be used to design filters. 
Starting with an initial condition as in (c), with two types of particles, characterized by $\gamma_A=5$ and  $\gamma_B=1$, 
we quickly arrive to a steady state where $A$ particles are confined to the left half of the box, while $B$ particles diffusive freely over the system. 
The lower panels in (c) and (d) indicate the density projected on the x-axis of $A$ and $B$ particles. 
}
\label{fig:snapshots}
\end{center}
\end{figure*}

We address this fundamental problem by using a simple model in which the active organisms move at 
constant speed in an heterogeneous two-dimensional space, where the heterogeneity is given by a random distribution of obstacles.  
An ``obstacle" may represent  the source of a  repellent chemical, a light gradient, a burning spot in a forest, or   
whatever threat that makes our (self-propelled) organisms to move away from it once the danger has been sensed; 
with obstacle avoidance characterized by a (maximum) turning speed $\gamma$. 
Our analysis shows that the same evolution equations (behavioral rules) lead to very different locomotion patterns at low and high density of obstacles. 
In the dilute obstacle scenario, there is no conflicting information and organisms can easily move away from the undesirable area they find in their way. On the other hand, when  we stress the environmental conditions, such that organisms sense several repellent sources simultaneously,  the processing of the information is no longer simple. 
Organisms compute the local obstacle density gradient and use this information to move away from higher obstacle densities. 
Since the distribution of obstacles is random, as the overall obstacle density increases, this task becomes increasingly more difficult. 
As result of this, no strategy  guaranties how to escape away from obstacles and the organisms behave more and more as if  there were no obstacles in the system. 
%
%
For low $\gamma$ values, we find that  the above described change of behavior is reflected by the minimum exhibited by the diffusion coefficient at  intermediate obstacle densities $\rho_o$, 
Fig.~\ref{fig:diffusion}(a).  
For large $\gamma$ values, particle motion is diffusive at small densities $\rho_o$, while for large  enough densities a new phenomenon emerges:  spontaneous trapping of particles, Fig.~\ref{fig:diffusion}(b).  
%
These traps are closed  orbits found by the particles in a landscape of obstacles, Fig.~\ref{fig:snapshots}. 
 %
 The time particles spend in these orbits is heavy-tailed distributed, and particle motion is genuinely (i.e., asymptotically) subdiffusive.  
 The boundary between the diffusive and subdiffusive regime depends on $\gamma$ and $\rho_o$ as illustrated in Fig.~\ref{fig:diffusion}(c).

 Our results open a new route to control  active particle systems. For instance, 
  the appearance of  spontaneous trapping as a dynamical phenomenon that depends on the intrinsic properties of the particles 
  allows us to design a generic filter of active particles. 

{\it Model definition.--}
We consider a continuum time model for $N_p$ self-propelled particles moving in a two-dimensional space of linear size $L$ where 
$N_o$ obstacles are placed at random~\cite{commentUS}. 
%
Boundary conditions are periodic. 
In the over-damped limit, the equations of motion of the $i$-th 
particle are given by:
\begin{eqnarray}
\label{eq:evol_x}
\dot{\mathbf{x}}_i &=& v_0 \mathbf{V}(\theta_i) \\
\label{eq:evol_theta}
\dot{\theta}_i       &=&   h(\mathbf{x}_i) +  \eta  \xi_{i}(t)  \, , 
\end{eqnarray}
%
where the dot denotes temporal derivative, $\mathbf{x}_i$ corresponds to the position of the $i$-th particle, and $\theta_i$ to its moving direction. 
The function $h(\mathbf{x}_i)$ represents the interaction with obstacles and its definition is given by: 
\begin{eqnarray}
\label{eq:h}
h(\mathbf{x}_i) = 
  \begin{cases}
   \frac{ \gamma}{n(\mathbf{x}_i)}  \sum_{\Omega_i} \sin(\alpha_{k,i} - \theta_i)  & \text{if } n(\mathbf{x}_i) > 0 \\
   0       & \text{if } n(\mathbf{x}_i) = 0 \, , 
  \end{cases}
\end{eqnarray}
where the sum runs over all neighboring obstacles $\Omega_i$ such that 
$0< |\mathbf{x}_i-\mathbf{y}_k|<R$, with $\mathbf{y}_k$ the position of the $k$-th obstacle, and the term $\alpha_{k,i}$ the angle, in polar coordinates, of the vector $\mathbf{x}_i - \mathbf{y}_k$. 
The term $n(\mathbf{x}_i)$ denotes the number of obstacles  located at a distance less or equal than $R$ from $\mathbf{x}_i$. 
%
%
In Eq.~(\ref{eq:evol_x}),  $v_{\rm 0}$ is the active particle speed and  $\mathbf{V}(\theta)\equiv (\cos(\theta),\sin(\theta))^T$.   
The additive white noise in Eq.~(\ref{eq:evol_theta}) is characterized by an amplitude $\eta$ and obeys  
 $\langle \xi_{i}(t) \rangle = 0$ and $\langle \xi_{i}(t) \xi_{j}(t') \rangle =\delta_{i,j} \delta(t-t')$, which leads to an angular diffusion $D_{\theta}=\eta^2/2$. 
%
%
%
Notice that for $\gamma=0$, equations~(\ref{eq:evol_x}) and~(\ref{eq:evol_theta}) define a system of  persistent random walkers characterized by a 
diffusion coefficient $D_{x_o}= v_0^2/(2D_{\theta})$, see~\cite{berg, okubo, peruani2007}.   
%
%


{\it Continuum description.--} 
We look for a  coarse-grain description of the system in terms of the concentration $p(\mathbf{x}, \theta, t)$ of particles at position $\mathbf{x}$ and orientation $\theta$ at time $t$.  
The evolution of $p(\mathbf{x}, \theta, t)$ obeys~\cite{gardiner}: 
\begin{eqnarray}
\label{eq:full_FP}
\partial_t p  + v_0 \nabla . \left[ \mathbf{V}(\theta) p \right]  = D_{\theta} \partial_{\theta \theta}p + F[p(\mathbf{x},\theta, t), \rho_o(\mathbf{x})]  \, ,
\end{eqnarray}
where $D_{\theta}$ is the angular diffusion as defined above, and $F[p(\mathbf{x},\theta, t), \rho_o(x)] $ represents the interaction of the self-propelled 
particles with the obstacles. 
The term $\rho_o(\mathbf{x})$ refers to the obstacle density at position $\mathbf{x}$~\cite{comment0}.
%
Here,  we discuss  two clear limits where $F[p(\mathbf{x},\theta, t), \rho_o(x)] $ can be specified.  We refer to these limits  as the low-density (LD) and high-density (HD) (obstacle) approximation.  

{\it Low-density approximation.--}
We consider that the active particles move most of the time freely,  bumping  into obstacles only occasionally. 
More specifically, we assume that $\eta, \rho_o<<1$ and approximate the interaction with obstacles, for time-scales much larger than $2R/v_0$, as sudden changes in the moving direction of the particle.
Let  $T(\theta, \theta'; \mathbf{x})$ be the rate at which a particle at position $\mathbf{x}$ and moving in direction $\theta$ turns  
into direction $\theta'$. 
To compute $T(\theta, \theta'; \mathbf{x})$ we need to estimate the frequency at which particles encounter  obstacles as well as the scattered angle 
after each obstacle interaction.  
If $D_{\theta}^{-1} v_0 >> \rho_o^{-1/2}$, we can approximate particle motion, in between successive encounters with obstacles, as ballistic.
In this limit, the obstacle encounter rate can be estimated as $\lambda(\rho_o) \approx v_o \rho_o \sigma_o$, where  $\sigma_o=2R$ is the associated scattering cross section.  
The absence of the classical constants of motion such as  angular momentum and mechanical energy prevents us from deriving an effective potential 
formalism from which to estimate the scattered angle. 
%
%
%
%
To simplify the calculations we approximate the scattered angle distribution by a simple top-hat functional form. 
Putting all this together, we express $T(\theta, \theta'; \mathbf{x}) \simeq \lambda(\rho_o) T(\theta, \theta') \approx \left[ \lambda(\rho_o)/(2 \epsilon_{\theta}) \right]  \Theta(\epsilon_{\theta} - |\theta - \theta' |)$ 
and express $F$ as: 
\begin{eqnarray}
\nonumber
F[{p}] &=& - \lambda(\rho_o) \, p(x,\theta, t) + \int_0^{2 \pi} d{\theta'}  T(\theta, \theta') p(x,\theta', t)  \\
\label{eq:F_dilute}
 &\approx &  \frac{\lambda(\rho_o) \epsilon_{\theta}^2}{6} \partial_{\theta \theta} p \, ,
\end{eqnarray}
where $\epsilon_{\theta}$ is numerically obtained from the study of the scattering process.  
Expression (\ref{eq:F_dilute}) allows us to rewrite the r.h.s. of  Eq.~(\ref{eq:full_FP})  as $\tilde{\mathcal{D}}_{\theta} \partial_{\theta \theta} p$, where $\tilde{\mathcal{D}}_{\theta}$ 
is defined as $\tilde{\mathcal{D}}_{\theta} = D_{\theta} + \lambda(\rho_o) \epsilon_{\theta}^2 / 6$. 
%
%
%
%
%
By performing a moment expansion of Eq.~(\ref{eq:full_FP}), where we define $\rho(\mathbf{x},t) = \int d\theta p$, $P_x(\mathbf{x},t) = \int d\theta \cos(\theta) p$, $P_y(\mathbf{x},t) = \int d\theta \sin(\theta) p$,  and  $Q_s(\mathbf{x},t) = \int d\theta \sin(2\theta) p$, and $Q_c(\mathbf{x},t) = \int d\theta \cos(2\theta) p$, we arrive to the following set of equations:
\begin{eqnarray}
\label{eq:rho}
\partial_t \rho &=& - v_0 \nabla.\mathbf{P} \\
\label{eq:Px}
\partial_t P_x &=&  - \frac{v_0}{2} \nabla . \left[ Q_c + \rho, Q_s \right] - \tilde{\mathcal{D}}_{\theta} P_x \\
\label{eq:Py}
\partial_t P_y &=& - \frac{v_0}{2} \nabla . \left[ Q_s,   \rho - Q_c \right] - \tilde{\mathcal{D}}_{\theta} P_y \,, 
\end{eqnarray}
%
where we assumed that   $\partial_t Q_c=\partial_t Q_s=0$. 
It  can be shown that the temporal evolution of $Q_c$ and $Q_s$ is faster than the one of $P_x$ and $P_y$, which in turn is faster than the one for $\rho$. 
Since we are interested in the long time behavior of $\rho(\mathbf{x},t)$, and there is no induced order, we take $Q_c=Q_s=0$ 
and use the fast relaxation of Eqs.~(\ref{eq:Px}) and~(\ref{eq:Py}) to express $P_x$ and $P_y$ as slave functions of $\rho$ and its derivatives~~\cite{comment1}. This procedure leads to the following asymptotic equation for $\rho(\mathbf{x},t)$:
\begin{eqnarray}
\label{eq:rho_asymptotic}
\partial_t \rho &=&  \nabla. \left[ \frac{v_0^2}{2 \tilde{\mathcal{D}}_\theta} \nabla \rho \right] \, .
\end{eqnarray}
From Eq.~(\ref{eq:rho_asymptotic}), it is evident that the spatial diffusion coefficient $D_x$ takes the form: $D_x = v_0^2/\left[2  \left( D_{\theta} + \Lambda_0 \rho_o  \right) \right]$, with $\Lambda_0 = v_0 \sigma_0 \epsilon_{\theta}^2 /6$. Notice that $D_x$ is a decreasing function of $\rho_o$~\cite{comment2}. 

\begin{figure}
\begin{center}
\resizebox{\columnwidth}{!} {\includegraphics{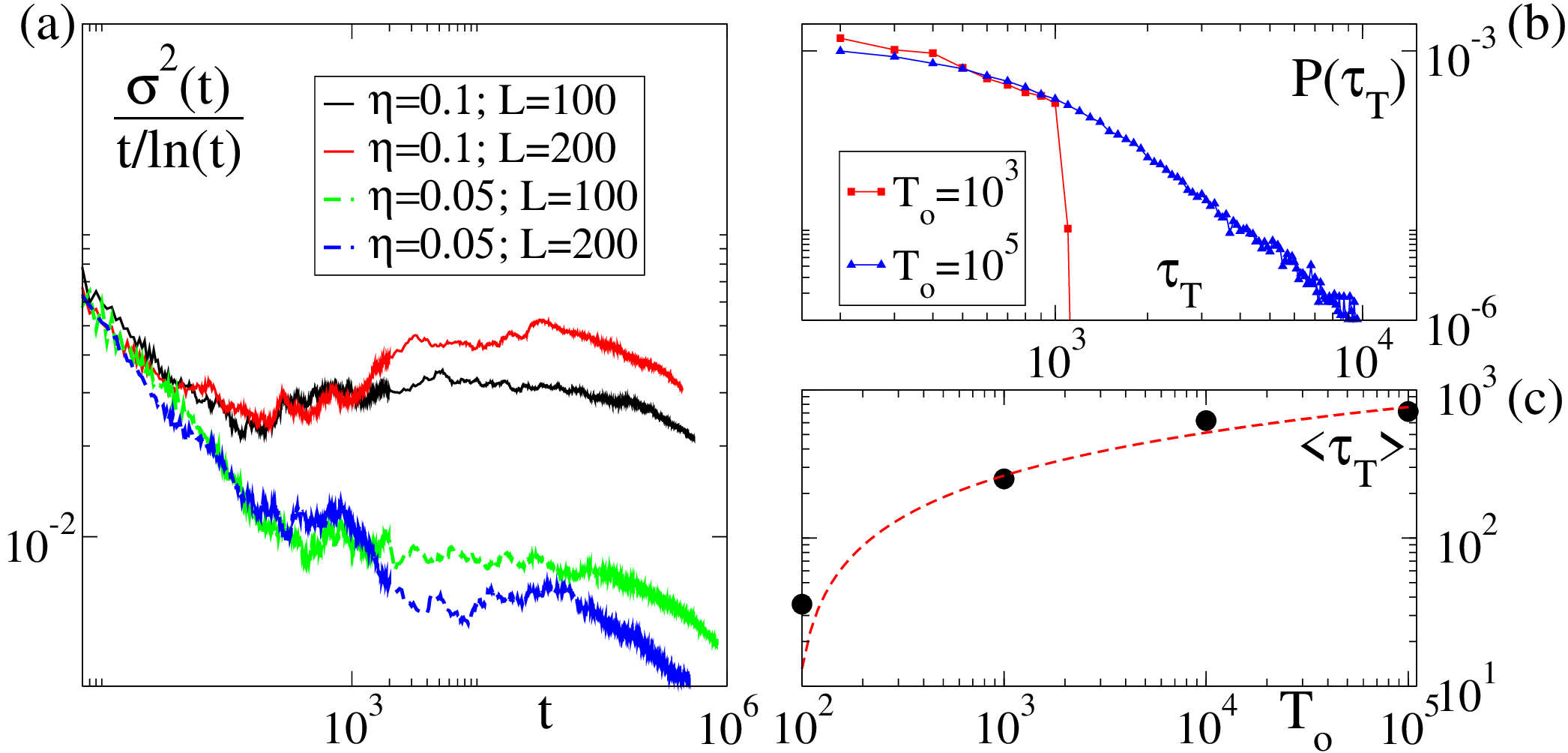}}
\caption{Genuine subdiffusive behavior. (a) scaling of the mean square displacement $\sigma^2(t)$ with $t$ for two system sizes, with $N_p/L^2=1$, $\rho_o=0.5$ and $\gamma=5$. Notice that the growth of $\sigma^2(t)$ is even slower than $t/\ln(t)$. 
(b) the distribution of trapping times $P(\tau_T)$ is power-law distributed for long enough $T_o$.  
(c) the average waiting time $\langle \tau_T \rangle$ is an increasing function of $T_o$. The red dashed curve corresponds to a fit $\propto \ln(T_o)$.  Measurements in (b) and (c) were performed on a particular trap for $\eta=0.1$.
}
\label{fig:subdiffusion}
\end{center}
\end{figure}

{\it High-density approximation.--} 
At large obstacle densities, particles always sense the presence of  several obstacles around them.   
%
This means that we cannot think of collisions as rare sudden jumps in the moving direction. 
Thus, we replace Eq.~(\ref{eq:F_dilute}) with a direct, local, coarse-grained expression for the interactions. 
This means that we leave the Boltzmann-like for the Fokker-Planck approach where the interaction with obstacles is expressed by 
$F = \partial_{\theta} \left[ I p(\mathbf{x}, \theta, t) \right]$. 
The term $I$ represents the (average) interaction felt by a particle at $\mathbf{x}$ moving in direction $\theta$, which takes the form: 
\begin{eqnarray}
\label{eq:F_highDensity}
I       =      \frac{\gamma}{n(\mathbf{x})} \sum_{j} \sin(\theta - \alpha_j) 
=   \frac{\gamma \Gamma(\mathbf{x}) }{n(\mathbf{x})} \sin(\theta - \psi(\mathbf{x})) \, ,
\end{eqnarray}
where $j$ is an index that runs over all neighboring obstacles, $\alpha_j$ is the  polar angle associated to the vector $\mathbf{x}-\mathbf{y}_j$, and 
$\Gamma(\mathbf{x})$  and $\psi(\mathbf{x})$ are the modulus and phase, respectively, of $\sum_j  \exp(\imath (\alpha_j))$, see~\cite{kuramoto1975}. 
%
%
We now approximate Eq.~(\ref{eq:F_highDensity}) by its average and use the fact that it represents a sum of $n$ random vectors of magnitude $1$ in the complex plane,  
to express $I \sim \sin(\theta - \psi(\mathbf{x}))/\sqrt{n}$, where $n \approx \pi R^2 \rho_o$.  
 Inserting this approximated expression into Eq.~(\ref{eq:full_FP}) and performing the moment expansion and approximations that led us from Eqs.~(\ref{eq:rho})-(\ref{eq:Py}) to  
 Eq.~(\ref{eq:rho_asymptotic}), we arrive to:
 \begin{eqnarray}
\nonumber
\partial_t \rho &=&  \frac{v_0^2}{2 D_{\theta}} \nabla^2 \rho - \frac{\gamma v_0}{2 D_{\theta}R\sqrt{\pi \rho_0}}  \nabla . \left[ (\cos(\psi),    \sin(\psi)) \rho   \right] \\
\label{eq:rho_asymptotic_HD}
&=&  \frac{v_0^2}{2 D_{\theta}} \nabla^2 \rho - \frac{\gamma v_0}{2 D_{\theta}R\sqrt{\pi \rho_0}}  \nabla . \left[ \frac{\rho \nabla \rho_o(\mathbf{x})}{|| \nabla \rho_o(\mathbf{x}) || }  \right] \, ,  
\end{eqnarray}
where we have approximated the vector field  $(\cos(\psi),    \sin(\psi))  \sim \nabla \rho_o(\mathbf{x})/ ||  \nabla  \rho_o(\mathbf{x})||$.
%
If we replace our current  definition of $\rho$ by a local average over a volume of linear dimensions 
much larger than $R$ and look for the long-time dynamics of this redefined density, by applying homogenization techniques, we expect to  recover  a diffusive behavior with a new effective diffusion,  whose explicit form depends on the statistical properties of the random field $\rho_o(\mathbf{x})$
 and is proportional to the square of the constant in front  of the convective term.  
While according to Eq.~(\ref{eq:rho_asymptotic}) (LD approx.),  $D_x \to 0$ as $\rho_o \to \infty$,  
Eq.~(\ref{eq:rho_asymptotic_HD}) (i.e., the HD approx.)  indicates that in the limit of $\rho_o \to \infty$,  $D_x \to D_{x0}$, where  $D_{x0}$ is the diffusion coefficient in absence of obstacles defined above. 
These two results necessarily imply the existence of a minimum in the spatial diffusion coefficient $D_{x}$ as $\rho_o$ is increased from $0$ to $\infty$. 
Moreover, this minimum has to be located at the crossover between the LD and HD approximation, which can be roughly estimated to occur at $\rho_c \sim 1/(\pi R^2)$, 
around which, $D_{x} \sim 1/\left[ \rho_c - \Lambda_1 \rho_o\right]$, with $\Lambda_1$ a constant.    
All these findings are confirmed by particle simulations as shown in Fig.~(\ref{fig:diffusion}).

{\it Trapping.--} 
Eq.~(\ref{eq:rho_asymptotic_HD}) indicates that the vector field  $\nabla \rho_o(\mathbf{x})$ governs the long-term dynamics at high obstacle concentrations.    
In particular, the random distribution of obstacles, together with the compressible nature of $\rho(\mathbf{x},t)$, may result in the spontaneous formation of active particle sinks. 
These topological defects, which we refer to as ``traps",  are indeed observed in simulations for large values of $\gamma$, Fig.~\ref{fig:snapshots}(a) and (b). 
%
%
Inside traps particles form vortex-like patterns. 
%
%
%
The average time $\langle \tau_T \rangle$ spent by a particle inside a trap depends on the precise configuration of the obstacles that form the trap. 
%
%
%
%
We find that the presence of  traps can lead to a genuine subdiffusive behavior, with particles exhibiting a mean-square displacement $\sigma^2(t)=\langle \mathbf{x}^2(t) \rangle$ that grows slower than  $t$. 
To test this observation, let us assume that particle motion can be conceived as a  random walk across a two dimensional array of traps such that  $\sigma^2(t) \propto n_J(t)$ where $n_J(t)$ represents the number of jumps from trap to trap the random walker performs during $t$. 
To  estimate $n_J(t)$, we study the distribution to trapping times $P(\tau_T)$ of a given trap, see Fig.~\ref{fig:subdiffusion}(b). 
Subdiffusion can only occur if $P(\tau_T)$ is asymptotically power-law distributed and such that 
 $\langle \tau_T \rangle$  grows with the observation time $T_o$ as $\ln(T_o)$ or faster~\cite{redner}, see Fig.~\ref{fig:subdiffusion}(b) and (c). 
 Within this simplified picture, the behavior of $\langle \tau_T \rangle$ shown in Figs.~\ref{fig:subdiffusion}(c) suggests that $\sigma^2(t) \propto n_J(t)  \sim t/\ln(t)$. 
 By taking $t \to \infty$, we expect $\sigma^2(t)/[t/\ln(t)]$ to approach a constant value. 
 Fig.~\ref{fig:subdiffusion}(a) clearly shows that $\sigma^2(t)$ is  slower than $t/\ln(t)$. 
 Arguably, this is due to  the fact that once a particle escapes from a trap, typically it performs a small excursion  before being reabsorbed by the same trap. 
 %
 

{\it Discussion.--} 
The spontaneous trapping of particles depends not only on $\rho_o$ but also on $\gamma$, which is an intrinsic property of the particle.  
This means that given the same spatial environment, two particle types, say characterized by $\gamma_A$ and $\gamma_B$, will respond differently. 
We can make use of this fact to fabricate a simple and cheap filter as indicated in Fig.~\ref{fig:snapshots}(c) and (d). 
Notice that trapping of one of the particle types is required to obtain this effect. 
A different diffusion coefficient for  $A$ and $B$ particles on the left half of the box does not suffice to induce higher concentration of one particle type to the left.
%

Trapping, rectification, and sorting have been reported for a particular kind of active particles: chiral, i.e. circularly moving particles~\cite{mijalkov2013, reichhardt2013}. 
By placing L-shaped obstacles on a regular lattice, the motion of such particles can be rectified~\cite{reichhardt2013}, while  
elongated obstacles arranged in flower-like patterns 
can be used to selectively trap either levogyre or dextrogyre particles~\cite{mijalkov2013}.   
For non-chiral active particle, trapping and rectification can be achieved by using V-shape objects. 
Kaiser et al. in~\cite{kaiser2012, kaiser2013} showed that self-propelled rods can be trapped by placing  V-shape objects. These traps 
provide a geometrical constrain to the active particles that end up being blocked in the V-shape devices. 
On the other hand, by arranging in line V-shape objects, but with their tips open, rectification of particle motion can be achieved~\cite{galajda2007, tailleur2009}. 
These V-shape objects, either with their tip closed or opened, cannot be used to produce a filter of active particles. 
Closed V-shape objects collect, by imposing geometric obstruction, any kind of self-propelled particle, while 
 in opened V-shape objects clogging of large size particles necessary occurs, preventing particle flow. 
Notice that the novel trapping and sorting mechanism reported here, which is based on the obstacle avoidance response time, is generic and should apply to  all kind of active particles, including interacting, non-interacting, chiral or non-chiral active particles.

 Finally, it is important to mention that  genuine subdiffusion  occurs for fixed obstacles only. 
 For slowly diffusing obstacles, i.e. for a (slow) dynamic environment, the asymptotic behavior of the active particles is diffusive. 
 The low and high density approximations, given by Eq.~(\ref{eq:rho_asymptotic}) and~(\ref{eq:rho_asymptotic_HD}), provide a reasonable description of particle motion even for dynamical environments as long as obstacle diffusion remains substantially smaller than the active motion. 
 Furthermore, trapping and subdiffusive behavior is also observed for interacting active particles as those studied in~\cite{chepizhko2013}  
 for values of the interaction strength significantly smaller than those associated to obstacle avoidance~\cite{commentMovies}. 

 Numerical simulations have been performed at the `Mesocentre SIGAMM' machine, hosted by Observatoire de la C{\^o}te d'Azur. 
 We thank F. Delarue and R. Soto for valuable comments on the manuscript and the Fed. Doeblin for partial financial support.

%
%
%
%
%
%
%
%
%
%
%
%
%
%
%

\bibliographystyle{apsrev}

\begin{thebibliography}{99}
\bibitem{boundaryEstimate} The boundary has been numerically estimated by: i) Studying $D_x(t\to \infty) = \langle \mathbf{x}^2(t)\rangle / 4t$ -- for subdiffusive particles $D_x(t)$ never saturates. 
ii) By measuring the asymptotic fraction of particle on the left half of the box in systems prepared as indicated in Fig.~\ref{fig:snapshots}(c) and (d) -- values of this quantity above $0.5$ can only occur due to the presence of traps. Both methods, i) and ii), led to the same boundary estimate.
\bibitem{comment0} We reserve the symbol $\rho_o$, without $\mathbf{x}$ dependency, to the global obstacle density.
\bibitem{comment1} Alternatively, multi-scaling analysis can be used to go from Eqs. (\ref{eq:full_FP}) and (\ref{eq:F_dilute}) to Eq. (\ref{eq:rho_asymptotic}).
\bibitem{comment2} There are qualitative differences with the Lorentz gas model, see~\cite{redner}. Firstly, in our model angular diffusion acts on the moving particles in between collisions. Secondly, the scattering process is not an instantaneous process and leads, given the absence of conserved quantities,  to a non bijective relationship between the impact parameter and the scattered angle.
\end{thebibliography}

\end{document}